\documentclass{elsart}

\usepackage{graphicx}
\graphicspath{{./img/}}
\usepackage{amsmath}
\usepackage{amssymb}
\begin{document}

\begin{frontmatter}

\title{Charge transport through a molecule driven by a high-frequency 
field\protect\footnote{This work is dedicated to Uli Weiss on the occasion 
of his 60th birthday.}}
\author{Sigmund Kohler},
\author{S\'ebastien Camalet},
\author{Michael Strass},
\author{J\"org Lehmann},
\author{Gert-Ludwig Ingold}, and
\author{Peter H\"anggi}
\address{Institut f\"ur Physik, Universit\"at Augsburg,
Universit\"atsstra\ss e~1, D-86135 Augsburg, Germany}

\begin{abstract}
  We study the current and the associated noise for the transport
  through a two-site molecule driven by an external oscillating field.
  Within a high-frequency approximation, the time-dependent
  Hamiltonian is mapped to a static one with effective parameters that
  depend on the driving amplitude and frequency. This analysis allows
  an intuitive physical picture explaining the nontrivial structure
  found in the noise properties as a function of the driving
  amplitude. The presence of dips in the Fano factor permits a control
  of the noise level by means of an appropriate external driving.
\end{abstract}

\end{frontmatter}

\section{Introduction}
\label{sec:introduction}

The development of techniques to contact molecules and to drive a current 
through them \cite{Reed1997a,Cui2001a,Reichert2002a} has opened the new 
and promising field of molecular electronics~\cite{Hanggi2002a,Nitzan2003a}.
In order to construct useful devices, however, it is not sufficient to
have a current flowing through a molecule but one also needs to have the 
ability to control this current. This can be achieved in principle by means
of the so-called single electron transistor setup where a gate electrode
is placed close to the molecule. Applying a gate voltage then allows to
influence the current across the molecule. In more complex circuits, the
need for a large number of contacts or electrodes close to the molecule may
constitute a major obstacle. In fact, already the implementation of a single
gate electrode which creates a sufficiently strong field at the molecule
is a demanding task~\cite{Liang2002a,Zhitenev2002a,Lee2003a}. 
Therefore, other means of controlling the current through a nanosystem should 
be explored. One possibility is to replace the static field of a gate 
electrode by a suitable external ac field. Recent theoretical 
work~\cite{Lehmann2003a} has demonstrated that, by using a 
coherent monochromatic field, one should indeed be able to control the 
electrical current flowing through a nanosystem connected to several leads. 
In an extension \cite{Camalet2003a} of this work, it was demonstrated that
even the noise level can be suppressed by an appropriate driving field.

In the present paper, we represent the molecule by a two-site system 
under the influence of an external high-frequency field and coupled to leads. 
Such a model is not limited to describe electrical transport through 
molecules but may also be applied to other situations like coherently
coupled quantum dot systems \cite{Blick1996a} irradiated by microwaves.

In Ref.~\cite{Camalet2003a}, a Floquet approach was
employed to derive exact expressions for both, the current and the
associated noise for the transport through a non-interacting
nanosystem in the presence of an arbitrary time-periodic field. A
study of the Fano factor, i.e. the ratio between noise and current,
revealed a suppression at certain values of driving amplitude and
frequency. To achieve a better physical understanding of this
phenomenon, we here consider the problem within the high-frequency
regime, which allows us to approximate the driven system by a static
one with renormalised parameters. The structure observed for the Fano
factor can then be understood in terms of three different scenarios.
For small effective intramolecular hopping matrix elements, the system 
itself acts as a bottleneck, while in the opposite limit, the two contacts 
form a two-barrier setup. In between, when the hopping matrix element is of
the order of the system-lead coupling strength, the barriers
effectively disappear, leading to a suppression of the Fano factor.

In the next section, we introduce our model, consisting of two sites
subject to an external oscillating field and coupled to two leads. We
then discuss this model in the static case, deriving explicit
expressions for both the current and its noise. In Sect.~\ref{sec:hf-approx}, 
expressions for the effective hopping matrix element and the electron 
distribution functions in the leads are determined within the high-frequency
approximation. These results are used in Sect.~\ref{sec:results} to compute 
current, noise and the corresponding Fano factor. A comparison with results 
based on the exact expressions of Refs.~\cite{Camalet2003a} 
demonstrates the validity of the high-frequency approximation for not too low
frequencies. This allows a physical interpretation of the observed features
in terms of a static model.

\section{The model}

\begin{figure}[t]
\centering
\includegraphics[width=0.60\columnwidth]{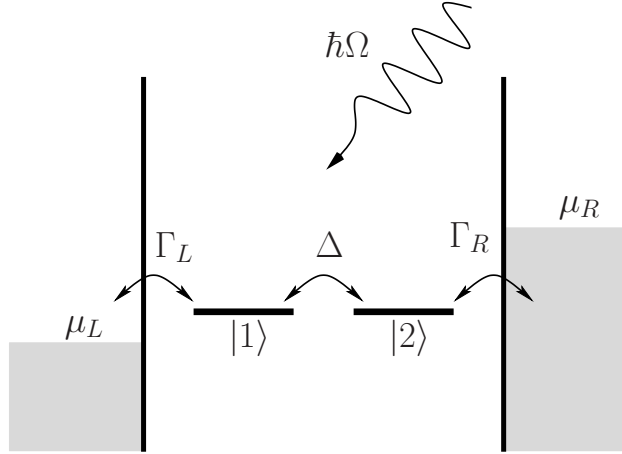}
\caption{Level structure of the nanoscale conductor with
two sites. Each site is coupled to the respective lead with chemical
potentials $\mu_L$ and $\mu_R=\mu_L+eV$.}
\label{fig:levels}
\end{figure}%
In the following, we consider the setup depicted in
Fig.~\ref{fig:levels}, which we describe by the time-dependent
Hamiltonian 
\begin{equation}
  H(t) = H_\mathrm{system}(t) + H_{\rm leads} + H_{\rm contacts}\ .
\end{equation}
The first term on the right-hand side,
\begin{equation}
  H_\mathrm{system}(t)
  = -\Delta(c_1^\dagger c_2+c_2^\dagger c_1)
  + \frac{A}{2}(c_1^\dagger c_1-c_2^\dagger c_2)\cos(\Omega t) ,
  \label{eq:HTLS}
\end{equation}
represents the driven two-site system, where electron-electron and
electron-phonon interactions have been disregarded. The fermion operators
$c_n$ and $c_n^{\dag}$, $n=1,2$, annihilate and create, respectively, an
electron at site $n$. Both sites are coupled by a hopping matrix
element $\Delta$.  The applied ac field with frequency
$\Omega=2\pi/\mathcal{T}$ results in a dipole force given by the second
term in the Hamiltonian~\eqref{eq:HTLS}. The amplitude~$A$ is
proportional to the component of the electric field strength parallel
to the system axis. 

The electrons in the leads are described by the Hamiltonian
\begin{equation}
H_\mathrm{leads}=\sum_q (\epsilon_{Lq}\,c^{\dag}_{Lq}
c^{\phantom{\dag}}_{Lq} + \epsilon_{Rq}\, c^{\dag}_{Rq} c^{\phantom{\dag}}_{Rq}),
\end{equation}
where $c_{Lq}^{\dag}$ ($c_{Rq}^{\dag}$) creates a spinless electron in
the left (right) lead with momentum $q$. The electron distribution in
the leads is assumed to be grand canonical with inverse temperature 
$\beta=1/k_B T$ and electro-chemical potential $\mu_{L/R}$. An applied 
voltage~$V$ corresponds to $\mu_R-\mu_L=eV$, where $-e$ is the electron charge.

The tunnelling Hamiltonian
\begin{equation}
H_{\rm contacts} = \sum_{q} \left( V_{Lq} c^{\dag}_{Lq} c^{\phantom{\dag}}_1
+ V_{Rq} c^{\dag}_{Rq} c^{\phantom{\dag}}_2
\right) + \mathrm{h.c.}
\end{equation}
establishes the contact between the sites and the leads, as
sketched in Fig.~\ref{fig:levels}.  The system-lead coupling is specified
by the spectral density
\begin{equation}
\Gamma_{\ell}(E)=2\pi\sum_q |V_{\ell q}|^2 \delta(E-\epsilon_{q\ell}) 
\end{equation}
with $\ell=L,R$. Below, we shall assume
within a so-called wide-band limit that these spectral densities are
energy independent, $\Gamma_{\ell}(E)=\Gamma_\ell$.

\section{Transport through a static two-site system}
\label{sec:static}

We start by deriving expressions for current and noise for a static two-site
system coupled to two leads, setting $A=0$ in the Hamiltonian~(\ref{eq:HTLS}). 
Solving the Heisenberg equations of motion for the lead operators, we obtain 
\begin{equation}
\label{eq:c_lead(t)}
c_{Lq}(t)=c_{Lq}(t_0)\mathrm{e}^{-\mathrm{i}\epsilon_{Lq}(t-t_0)/\hbar}
-\frac{\mathrm{i}V_{Lq}}{\hbar}\!\!\int\limits_{t_0}^{t}\! \mathrm{d}t'\,
\mathrm{e}^{-\mathrm{i}\epsilon_{Lq}(t-t')/\hbar} c_1(t')
\end{equation}
and a corresponding expression for $c_{Rq}(t)$ with $L$ replaced by $R$
and $c_1$ by $c_2$. Inserting \eqref{eq:c_lead(t)} into the Heisenberg
equations of motion of the two-site system and exploiting the wide-band
limit, one arrives at
\begin{align}
\dot c_{1} =& \frac{\mathrm{i}}{\hbar} \Delta \, c_{2}
             -\frac{\Gamma_{L}}{2\hbar}c_{1} + \xi_{L}(t), 
\label{eq:c1}
             \\
\dot c_{2} =& \frac{\mathrm{i}}{\hbar} \Delta \, c_{1}
             -\frac{\Gamma_{R}}{2\hbar}c_{2} + \xi_{R}(t).
\label{eq:c}
\end{align}
For a grand canonical ensemble, the operator-valued
Gaussian noise
\begin{equation}
\xi_{\ell}(t)=-\frac{\mathrm{i}}{\hbar}\sum_q V^*_{\ell q}
\exp\left[-\frac{\mathrm{i}}{\hbar}\epsilon_{\ell q}(t-t_0)\right]
c_{\ell q}(t_0)
\end{equation}
obeys
\begin{align}
\label{xi}
\langle\xi_\ell(t)\rangle &= 0,
\\
\label{xi2}
\langle\xi^\dagger_\ell(t)\,\xi_{\ell'}(t')\rangle
&= \delta_{\ell\ell'}\frac{\Gamma_\ell}{2\pi\hbar^2}
\int\!\! \mathrm{d}\epsilon\, \mathrm{e}^{\mathrm{i}\epsilon(t-t')/\hbar}f_\ell(\epsilon),
\end{align}
where $f_\ell(\epsilon)=\left\{1+\exp[\beta(\epsilon-\mu_\ell)]\right\}^{-1}$
denotes the Fermi function with chemical potential $\mu_\ell$, $\ell=L,R$.
In the asymptotic limit $t_0\to -\infty$, the solutions of
Eqs.~\eqref{eq:c1} and \eqref{eq:c} read with $n=1,2$:
\begin{equation}
\label{eq:c(t)}
c_n(t)=\int\limits_0^\infty
 \mathrm{d}\tau\, \big\{ G_{n1}(\tau) \, \xi_L(t-\tau) 
+ G_{n2}(\tau) \, \xi_R(t-\tau)\big\}.
\end{equation}
In the wide-band limit and for equal system-lead coupling,
$\Gamma_{\ell}=\Gamma$, the propagator is given by
\begin{equation}
\label{eq:prop}
G(\tau) = \mathrm{e}^{-\Gamma\tau/2} 
\begin{pmatrix}
\cos(\Delta\tau) & \mathrm{i}\sin(\Delta\tau)\\
\mathrm{i}\sin(\Delta\tau) & \cos(\Delta\tau)
\end{pmatrix}\Theta(\tau),
\end{equation}
where $\Theta(\tau)$ is the Heaviside step function.

The operators corresponding to the currents across the contacts
$\ell=L,R$ are given by the negative time derivative
of the electron number $N_\ell=\sum_q c^{\dag}_{\ell q}
c^{\phantom{\dag}}_{\ell q}$ in
the respective lead, multiplied by the electron charge~$-e$.
For the current through the left contact one finds
\begin{equation}
\begin{split}
\label{eq:I(t)}
I_L(t)&=\frac{\mathrm{i}e}{\hbar}\sum_{q} \left( V_{Lq}^* c^{\dag}_1 
c^{\phantom{\dag}}_{Lq} - \mathrm{h.c.}\right)\\
&=\frac{e}{\hbar}\Gamma_Lc_1^\dagger(t)c_1(t)
       -e\big\{c_1^\dagger(t)\xi_L(t)+\xi_L^\dagger(t)c_1(t)\big\}
\end{split}
\end{equation}
with a corresponding expression for $I_R(t)$.  In the stationary limit, $t_0\to-\infty$,
the mean values of the currents across the two contacts agree and we
obtain
\begin{equation}
I = \langle I_L\rangle= \frac{e}{2\pi\hbar}\int\d E\, \big[f_R(E)-f_L(E)\big]T(E).
\label{eq:I}
\end{equation}
In the wide-band limit, the transmission $T(E)$ can be expressed in terms of
$G_{12}(E)$, i.e. the Fourier transform of the propagator $G_{12}(\tau)$, as
\begin{equation}
T(E) = \Gamma_L\Gamma_R |G_{12}(E)|^2 .
\end{equation}
Making use of the propagator \eqref{eq:prop}, the
transmission for $\Gamma_{\ell}=\Gamma$ becomes
\begin{equation}
T(E) = \frac{\Gamma^2\Delta^2}{|(E-\mathrm{i}\Gamma/2)^2-\Delta^2|^2} .
\label{eq:T}
\end{equation}

The noise of the current through contact $\ell$ is given by the symmetric 
auto-correlation function of the current fluctuation operator 
$\Delta I_\ell(t) = I_\ell(t)-\langle I_\ell(t)\rangle$.  
It is possible to characterise the noise strength by its zero frequency
component 
\begin{equation}
S= \frac{1}{2}\int_{-\infty}^{+\infty}\d t \langle\Delta I_\ell(t)
\Delta I_\ell(0) +\Delta I_\ell(0) \Delta I_\ell(t)\rangle,
\end{equation}
which is independent of the contact $\ell$. The quantity $S$ may be 
expressed in terms of the transmission function~$T(E)$ as 
\cite{Blanter2000a}
\begin{equation}
\begin{split}
S = \frac{e^2}{2\pi\hbar}\int\d E\,\Big\{ &
T(E) \big[f_L(E)[1-f_L(E)] + f_R(E)[1-f_R(E)] \big] \Big. \\
&+ T(E)\big[1-T(E)\big] \big[f_R(E)-f_L(E)\big]^2 \Big\}. \label{eq:sn}
\end{split}
\end{equation}
Two contributions to the zero-frequency noise $S$ have to be distinguished: 
The first term is a temperature-dependent equilibrium noise according to the
dissipation-fluctuation theorem \cite{Callen1951a} and dominates for
$\beta eV \ll 1$. In contrast, for large voltages $\beta eV\gg1$, the
main contribution to the noise stems from the second term. This so-called 
shot noise has its physical origin in the discreteness of the charge carriers.

We now consider voltages larger than all other energy scales in the
problem. As a consequence, the current noise will entirely be due to 
shot noise. Furthermore, in this limit, the results for current and 
noise will not depend on temperature.
In the expression (\ref{eq:I}) for the current, the difference of the
Fermi distributions then practically equals one for energies where the
transmission is nonvanishing. The current thus reads
\begin{equation}
I_{\infty} = \frac{e}{2\pi\hbar} T
= \frac{e\Gamma}{2\hbar}\, \frac{\Delta^2}{\Delta^2+(\Gamma/2)^2}\ ,
\label{eq:Iinf}
\end{equation}
where $T=\int\d E\,T(E)$ is the total transmission.
With the same argument we find from (\ref{eq:sn}) for the current noise
\begin{equation}
S_{\infty} = 
 \frac{e^2\Gamma}{\hbar}\,
\frac{2
  \Delta^2(\Gamma^4-2\Gamma^2\Delta^2+8\Delta^4)}{(4\Delta^2+\Gamma^2)^3}\ .
\label{eq:Sinf}
\end{equation}
The relative noise strength can be characterised by the so-called Fano
factor $F= S/eI$ which, in the infinite voltage limit, becomes
\begin{equation}
  \label{fano}
 F_{\infty}=
 \frac{\Gamma^4-2\Gamma^2\Delta^2+8\Delta^4}{(4\Delta^2+\Gamma^2)^2}\ .
\end{equation}

\begin{figure}[tbp]
  \centering
  \includegraphics{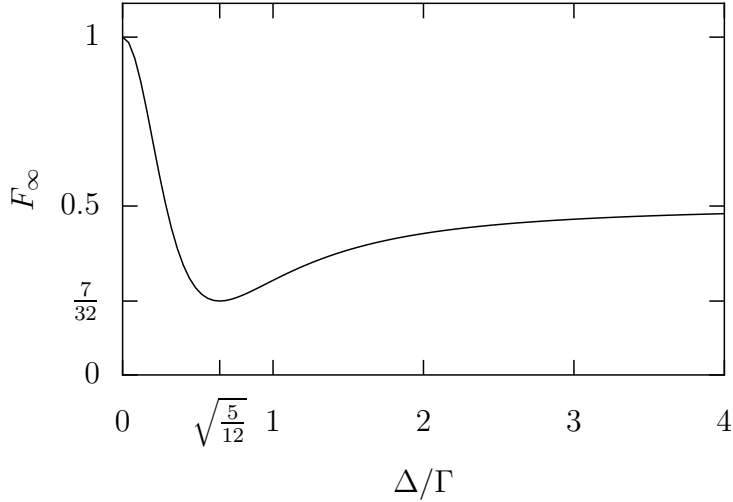}
  \caption{Fano factor~$F_{\infty}=S_{\infty}/eI_{\infty}$ as a function 
    of $\Delta/\Gamma$.
    For $\Delta\ll\Gamma$, the bottleneck of the transport is the
    tunnelling process between the two sites yielding a Fano factor
    $F_{\infty}=1$. In the opposite limit $\Delta\gg\Gamma$, we obtain
    transport through a double-barrier structure with a corresponding
    Fano factor $F_{\infty}=1/2$. In the intermediate regime 
    the Fano factor
    assumes a minimum at the position indicated in the plot.  }
  \label{fig:fanodelta}
\end{figure}%

In Fig.~\ref{fig:fanodelta}, the Fano factor $F_{\infty}$ is depicted as a 
function of the ratio of the tunnelling matrix element $\Delta$ and the 
level width $\Gamma$.  For weak system-lead coupling 
$\Gamma\ll\Delta$, the two contacts between the two-site system and
the leads form the limiting step of the transport process.  We
effectively arrive at transport through a double-barrier system with a
Fano factor $F_{\infty}=1/2$ \cite{Chen1991a}. 
On the other hand, for
$\Gamma\gg\Delta$ the two sites hybridise with the adjacent lead
and effectively only a single barrier remains. This yields a Fano
factor $F_{\infty}=1$. 
At the crossover between these two regimes, the channel
is optimally transparent and, consequently, the Fano factor assumes a
minimum. From the expression (\ref{fano}), we find the optimal
hopping matrix element $\Delta=\sqrt{5/12}\,\Gamma$ yielding a minimal
Fano factor of $F_{\infty}=7/32$. We remark that the minimum decreases 
further if the number of sites in the system is increased 
\cite{Camalet2003a}.

\section{High-frequency approximation}
\label{sec:hf-approx}

Let us now turn back to the original time-dependent problem.  We will compute
within a high-frequency approximation \cite{Grossmann1991b} the current 
through this driven system and the corresponding current noise. Results valid 
for arbitrary driving amplitudes $A$ can be obtained by the following procedure 
which is justified in the Appendix on the basis of Floquet theory.

First, we introduce the interaction picture with respect to the driving which 
for the problem at hand is obtained by means of the unitary transformation
\begin{equation}
\label{trafo}
U_0(t) = \exp\left(-\mathrm{i}\frac{A}{2\hbar\Omega}
(c_1^\dagger c_1-c_2^\dagger c_2)\sin(\Omega t)\right) .
\end{equation}
This yields the new system operators
\begin{equation}
\tilde c_{1,2}(t) = U_0^\dagger(t) c_{1,2} U_0(t)
= c_{1,2}\exp\left(\mp\mathrm{i}\frac{A}{2\hbar\Omega}\sin(\Omega t)\right) ,
\end{equation}
where the upper sign corresponds to site 1. To a good approximation, the 
dynamics can then be described by the time-averaged system Hamiltonian 
\begin{equation}
\bar H_\mathrm{system} 
= -\Delta_\mathrm{eff}(c_1^\dagger c_2+c_2^\dagger c_1) .
\end{equation}
Thus, within a high-frequency approximation, the driven two-site system acts
as a static system with the effective hopping matrix element
\begin{equation}
\label{eq:deltaeff}
\Delta_\mathrm{eff} = J_0(A/\hbar\Omega)\Delta ,
\end{equation}
where $J_0$ is the zeroth order Bessel function of the first kind.
The driving amplitude~$A$ and frequency~$\Omega$ can now be chosen
such that $\Delta_\mathrm{eff}$ vanishes and consequently tunnelling
between the two central sites then no longer
occurs~\cite{Grossmann1992a,Grossmann1991a,Grossmann1991b}.

Proceeding as in Sect.~\ref{sec:static}, the influence of the leads
after the transformation (\ref{trafo}) can be described by fluctuation 
operators. For the left lead one finds
\begin{equation}
\label{eta(t)}
\eta_L(t) = -\frac{\mathrm{i}}{\hbar}\sum_q V^*_{Lq}\exp\left[-
\frac{\mathrm{i}}{\hbar}\left(\epsilon_{Lq}(t-t_0)+
\frac{A}{2\Omega}\sin(\Omega t)\right)\right] c_{Lq}(t_0)
\end{equation}
with the correlation function
\begin{eqnarray}
\langle\eta_L^\dagger(t+\tau) \eta_L(t)\rangle
&=& \frac{\Gamma_L}{2\pi\hbar^2} \int\d\epsilon\sum_{k,k'}
    \e^{\mathrm{i}\epsilon\tau/\hbar}f_L(\epsilon+k\hbar\Omega)
\nonumber\\
& & \hspace{2truecm}\times J_k(A/2\hbar\Omega) J_{k'}(A/2\hbar\Omega)
    \e^{-\mathrm{i}(k-k')\Omega t}
\label{correlation:driven}
\end{eqnarray}
and corresponding expressions for the right lead.
Since we are interested in the average current and the zero-frequency
noise, i.e. low-frequency transport properties, we can neglect the
$\mathcal{T}$-periodic contribution to the correlation function
\eqref{correlation:driven} and, thus, average its $t$-dependence over
the driving period.  Then, the correlation function assumes the form
\eqref{xi2} like in the static case but with the Fermi function
replaced by the effective distribution function
\begin{equation}
\label{f:eff}
f_{\ell,\mathrm{eff}}(E)
= \sum_{k=-\infty}^{\infty} J_k^2(A/2\hbar\Omega) f_\ell(E + k\hbar\Omega) .
\end{equation}
The different terms in this sum describe processes where an electron of
energy~$E$ is transferred from lead~$\ell$ to the system under
absorption (emission) of $|k|$ photons for $k<0$ ($k>0$). These
processes are weighted by the square of the $k$th order Bessel
function of the first kind.  

Having approximated the originally time-dependent problem by a static
one with an effective hopping matrix element and an effective
distribution function, we can calculate the transmission, the current,
and the zero frequency noise of the \textit{driven} system with the
formulae derived in Sect.~\ref{sec:static} for a \textit{static}
situation.  

In the limit of very large voltages and for energies where the transmission
is nonvanishing, the effective distribution functions in the left and right
lead become again zero or one, respectively. As a consequence, the
time-averaged current and the zero-frequency noise are given by the
expressions (\ref{eq:Iinf}) and (\ref{eq:Sinf}) with
the replacement $\Delta \rightarrow \Delta_\mathrm{eff}$. We denote
the current and the noise in this limit by ${\bar I}_{\infty}$ and
$\bar S_\infty$, respectively. As pointed out above, there exist
driving parameters where the effective hopping matrix element
(\ref{eq:deltaeff}) vanishes. As a consequence, no current can
flow through the system under these 
circumstances~\cite{Lehmann2003a,Camalet2003a}.

\begin{figure}[t]
\centering
\includegraphics[width=0.60\columnwidth]{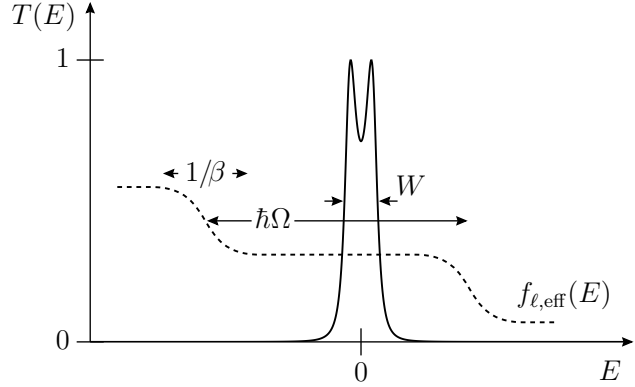}
\caption{Typical energy dependence of transmission $T(E)$ (solid line) and 
effective distribution function $f_{\ell,\mathrm{eff}}(E)$ (dashed line) 
which allows to replace the distribution function by (\protect\ref{f:eff2}) 
in the expressions for current and noise.}
\label{fig:peakdist}
\end{figure}

The case of a finite voltage requires a more detailed inspection of the
distribution functions $f_{\ell,\mathrm{eff}}$ and the effective transmission
$T(E)$ sketched in Fig.~\ref{fig:peakdist}. In the high-frequency regime 
under study here, the width $W$ of the transmission function is much
smaller than $\hbar\Omega$. Furthermore, the effective distribution function 
$f_{\ell,\mathrm{eff}}$ is nearly constant for energies $E$ separated by at 
least $1/\beta$ from the steps at $E=\mu_\ell+k\hbar\Omega$. Therefore, unless 
a step in $f_{\ell,\mathrm{eff}}$ occurs close to $E=0$, the effective 
distribution functions in the current and noise expressions can be replaced 
by their value at $E=0$, i.e.
\begin{equation}
\label{f:eff2}
f_{\ell,\mathrm{eff}} = \sum_{k<\mu_\ell/\hbar\Omega} J_k^2(A/2\hbar\Omega) .
\end{equation}

Thus, the time-averaged current and the zero-frequency noise are given by
\begin{align}
{\bar I} &= {\bar I}_{\infty} \Big( J_0^2(A/2\hbar\Omega) 
+ 2 \sum_{k=1}^{K(V)} J_k^2(A/2\hbar\Omega) \Big), \label{eq:Iapprox} \\
{\bar S} &= \frac{e}{2} {\bar I}_{\infty}+ \Big( J_0^2(A/2\hbar\Omega) 
+ 2 \sum_{k=1}^{K(V)} J_k^2(A/2\hbar\Omega) \Big)^2  
\Big( {\bar S}_{\infty} - \frac{e}{2}{\bar I}_{\infty} \Big) ,
\label{eq:Sapprox}
\end{align} 
where $K(V)$ denotes the largest integer not exceeding $eV/2\hbar\Omega$.
Note that the Fano factor $F={\bar S}/e{\bar I}$ for fixed $A/\Omega$ reaches 
its minimal value in the infinite voltage limit. 
Since $J_k(x)\approx 0$ for $x > k$ and $\sum_k J_k^2(x)=1$, the dc
current and the zero frequency noise are well approximated for $A<eV$
by their asymptotic values for infinite voltage, ${\bar I} \approx
{\bar I}_{\infty}$ and ${\bar S} \approx {\bar S}_{\infty}$. We remark
that, in contrast to the static case, the result~\eqref{eq:Sapprox}
contains contributions stemming from the first term in the noise expression
\eqref{eq:sn} even in the zero-temperature limit.

\section{Comparison with exact results}
\label{sec:results}

\begin{figure}
\centering
\includegraphics{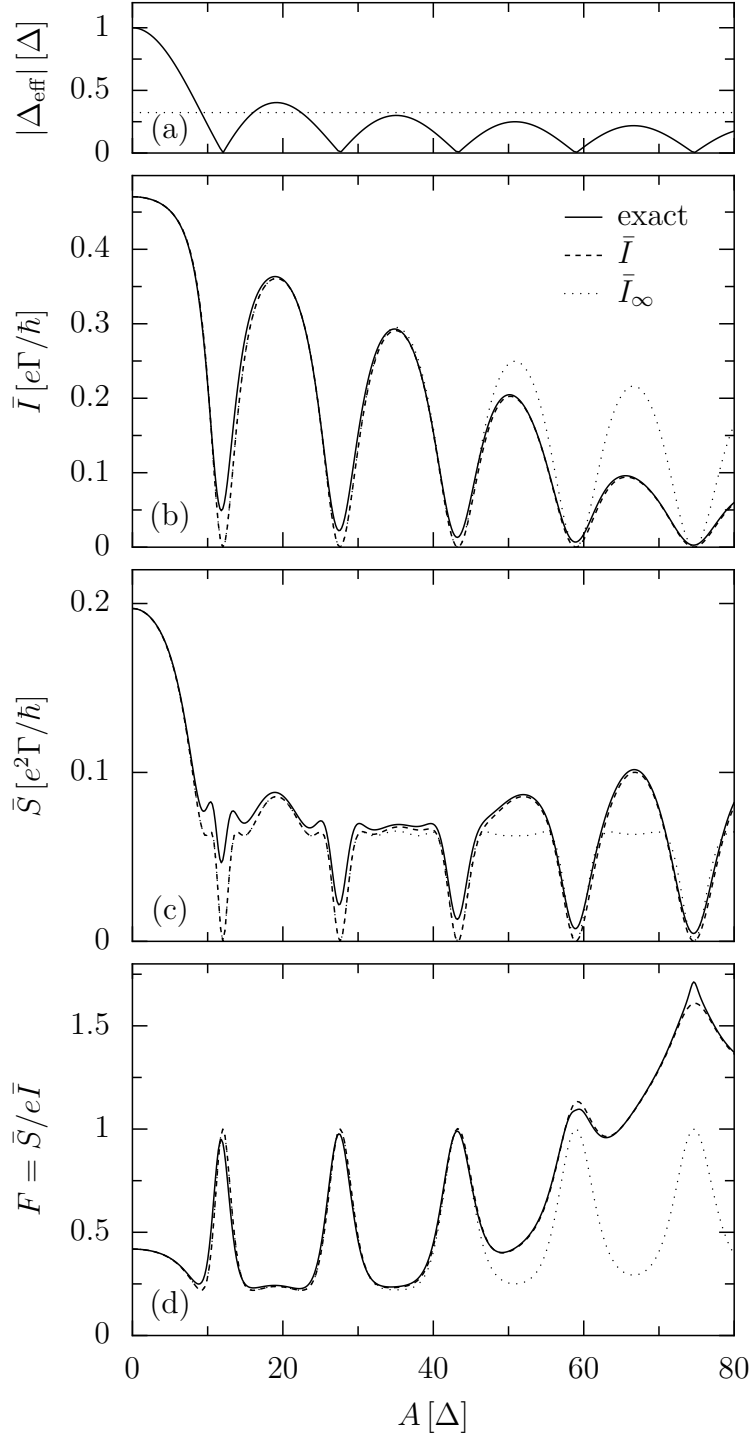}
\caption{(a) Effective hopping matrix element $|\Delta_\mathrm{eff}|$, (b)
  time-averaged current $\bar I$, (c) zero-frequency noise $\bar S$, and
  (d) Fano factor $F=\bar S/e \bar I$ as a function of the driving amplitude 
  $A$. Shown are the numerically exact results (solid lines), the
  approximative results~\eqref{eq:Iapprox} and \eqref{eq:Sapprox} for
  finite voltage (dashed lines), and the infinite voltage results
  \eqref{eq:Iinf} and \eqref{eq:Sinf} with $\Delta$ replaced by
  $\Delta_\mathrm{eff}$ (dotted lines). The coupling strength is
  $\Gamma=0.5\,\Delta$, the driving frequency is $\Omega=5\,\Delta/\hbar$,
  and the voltage reads $V=48\,\Delta/e$.  The dotted line in (a) marks the
  value $\sqrt{5/12}\,\Gamma$ for which the Fano factor assumes its minimum.}
\label{fig:lV}
\end{figure}%

Figures \ref{fig:lV}b--d depict by solid lines the time-averaged current, 
the zero-frequency noise and the Fano factor at zero temperature obtained
numerically within the Floquet approach of Ref.~\cite{Camalet2003a}
for the relatively large voltage $V=48\Delta/e$.  This particular
value of the voltage has been selected to avoid the chemical potentials
to lie close to multiples of $\hbar \Omega$. A comparison of these
numerically exact results for current and noise with the approximate 
expressions~\eqref{eq:Iapprox} and \eqref{eq:Sapprox} depicted 
by dashed lines shows a good agreement for the parameters chosen.  
The agreement improves with increasing frequency: already for 
$\Omega=10\Delta/\hbar$, it is found that the exact and approximate results 
can practically no longer be distinguished.

The exact numerical results show strong suppressions of both, the current and
the noise for certain driving amplitudes. This behavior can be explained
within the high-frequency approximation presented in
Section~\ref{sec:hf-approx}: Whenever the ratio $A/\hbar\Omega$ corresponds to
a zero of the Bessel function $J_0$, the effective hopping matrix element
$\Delta_\mathrm{eff}$ vanishes (cf.\ Fig.~\ref{fig:lV}a) and consequently the
current and the noise become zero.  Note that the exact result exhibits still
a residual current and noise.
The suppressions of the current and noise lead to peaks of
the Fano factor $F$.  For sufficiently small driving amplitudes, these
peaks are accompanied by minima which correspond to
$|\Delta_\mathrm{eff}|\simeq\sqrt{5/12}\,\Gamma$ indicated by the dotted line in
Fig.~\ref{fig:lV}a.

For driving amplitudes $A\lesssim eV$, the finite voltage results
\eqref{eq:Iapprox} and \eqref{eq:Sapprox} for the current $\bar I$ 
and the noise $\bar S$ are well described by the
results \eqref{eq:Iinf} and \eqref{eq:Sinf} for infinite voltage with
$\Delta$ replaced by $\Delta_\mathrm{eff}$.  In this regime, the Fano
factor reaches maxima $F=1$.  In contrast, for larger driving amplitudes
$A>eV$, we find a Fano factor larger than that predicted by
\eqref{fano}, as discussed below \eqref{eq:Sapprox}.  In particular, the
Fano factor can assume values $F>1$.

\begin{figure}
\centering
\includegraphics{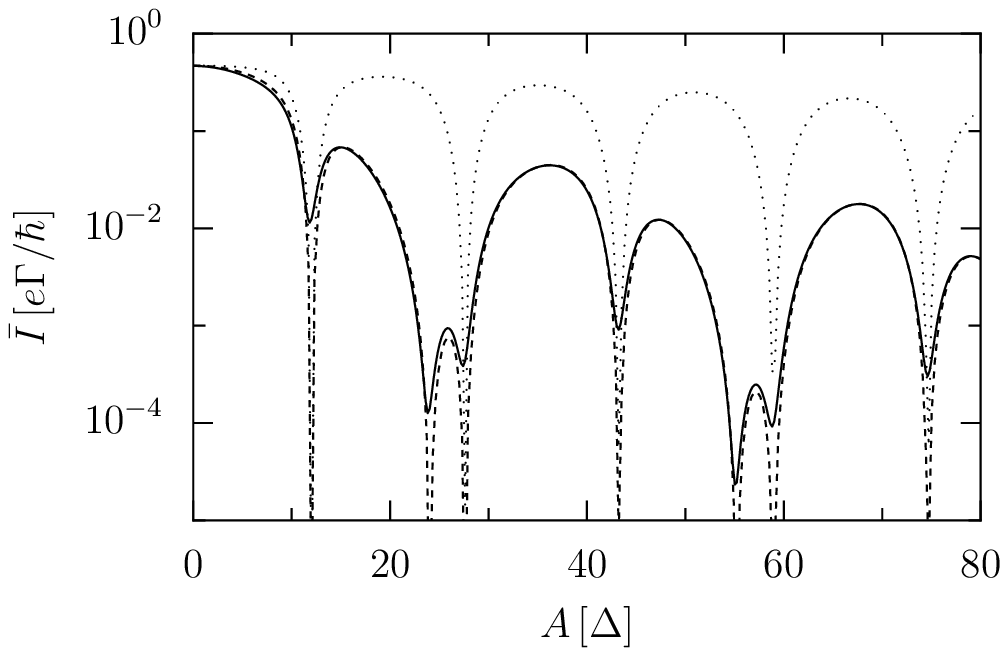}
\caption{Time-averaged current $\bar I$  as
  a function of the driving amplitude $A$. Shown are the numerically
  exact result (solid line), the approximate result~\eqref{eq:Iapprox}
  for finite voltage (dashed line) and the infinite voltage result (dotted
  line). The coupling strength is $\Gamma=0.5\,\Delta$, the driving
  frequency is $\Omega=5\,\Delta/\hbar$, and the voltage reads
  $V=5\,\Delta/e$.}
\label{fig:sV}
\end{figure}%
Finally, we consider in Fig.~\ref{fig:sV} the case of intermediate voltages 
such that $\Delta,\Gamma<eV<2\hbar\Omega$. Then, only the zero photon
channel contributes and hence the current
${\bar I}={\bar I}_{\infty} J_0^2(A/2\hbar\Omega)$ is considerably lower
than for large voltages.  Now, in addition, a new type of
suppression appears at twice the amplitude compared to the suppressions 
discussed above. The physical reason for this new kind of suppression lies
in the fact that the effective distribution functions in the two leads are
equal at the relevant energies and therefore no dc current can flow.
Nevertheless, the noise remains finite and, consequently, the Fano factor 
diverges.

\section{Conclusions}

We have presented a high-frequency approximation for the charge transport
through a driven two-site system.  Within this scheme, the time-dependent
Hamiltonian and the lead correlation functions are transformed to an
appropriate interaction picture and subsequently time-averaged over the
driving period.  In the resulting equations, the hopping matrix element of
the two-site system and the electron distributions of the attached leads
are replaced by effective ones which depend on the driving parameters.

This static picture allows to gain profound physical insight into the structure
present in current and noise as a function of the driving parameters.
For small effective hopping matrix elements, the barrier between the two
sites of the system dominates, leading to shot noise with $F=1$.
In the opposite limit, the contacts form a double-barrier system corresponding 
to a Fano factor $F=1/2$. Between these two situations, effectively no
barrier exists in the transport path and the Fano factor is further
reduced to a minimal value of $F=7/32$.

The results of this work demonstrate that the control of a current through 
a molecule by means of a time-periodic driving provides a viable alternative 
to the traditional single electron transistor setup based on a gate 
electrode. Our approach allows to minimise the number of electrodes close to
the molecule. Furthermore, a suitable choice of the transistor's
working point permits to operate in a low noise regime with a small Fano
factor. These features inherent to the driven setup may prove useful 
for the development of novel molecular electronics devices.

\section*{Acknowledgements}

The authors acknowledge financial support by a Marie Curie fellowship
of the European community program IHP under contract No.\ 
HPMF-CT-2001-01416 (S.C.), by the Volkswagen-Stiftung under Grant No.\ 
I/77 217, and by the DFG through Graduiertenkolleg~283 and
Sonderforschungsbereich~486.

\appendix
\section{Driven quantum systems in high-frequency approximation}
\label{app:approximation}

In this Appendix, we review a common perturbative approach for the
treatment of periodically time-dependent quantum systems and thereby
justify the high-frequency approximation employed in
Section~\ref{sec:hf-approx}.

A standard technique for the study of periodically time-dependent
Hamiltonians $H(t)=H(t+\mathcal{T})$ is the so-called Floquet approach
\cite{Shirley1965a,Sambe1973a,Grifoni1998a}.
It starts out from the fact that a complete set of solutions of the
corresponding Schr\"odinger equation is of the form $|\psi_\alpha(t)\rangle
= \e^{-\mathrm{i}\epsilon_\alpha t/\hbar}|\phi_\alpha(t)\rangle$ where the
Floquet states $|\phi_\alpha(t)\rangle=|\phi_\alpha(t+\mathcal{T})\rangle$
obey the time-periodicity of the Hamiltonian.  The Floquet states and the
quasi-energies $\epsilon_\alpha$ are eigenstates and eigenvalues,
respectively, of the Hermitian operator $\mathcal{H} =
H(t)-\mathrm{i}\hbar\mathrm{d}/\mathrm{d} t$ defined in a Hilbert space
extended by a periodic time coordinate.  We emphasise that already the
Floquet states from a single Brillouin zone $-\hbar\Omega/2 \leq 
\epsilon_\alpha < \hbar\Omega/2$ form a complete set of solutions.  A 
Floquet ansatz essentially maps the time-dependent problem to an eigenvalue 
problem and, thus, it is possible to employ all approximation schemes known 
from time-independent quantum mechanics, in particular perturbative 
schemes for the computation of eigenstates.

Here, we consider the special case of a time-dependent Hamiltonian of the 
form
\begin{equation}
\label{app:H}
H = H_0 f(t) + H_1
\end{equation}
where $f(t)$ is a $\mathcal{T}$-periodic function with zero
time-average.  If $\hbar\Omega$ is much
larger than all energy differences in the spectrum of $H_1$, the following
Schr\"odinger perturbation theory can be employed for the computation of the
Floquet states \cite{Shirley1965a,Sambe1973a}: It is assumed that for $H_0$
all eigenstates $|\varphi_\alpha\rangle$ and eigenenergies $E_\alpha$ are
known.  Then, the unperturbed Floquet Hamiltonian $\mathcal{H}_0=H_0
f(t)-\mathrm{i}\hbar\mathrm{d}/\mathrm{d} t$ has the eigensolutions
\begin{equation}
|\phi_\alpha^k(t)\rangle = \exp\left(-\frac{\mathrm{i}}{\hbar}E_\alpha F(t)+
\mathrm{i}k\Omega t\right)|\varphi_\alpha\rangle
\label{eq:basetrafo}
\end{equation}
with eigenvalue 
$k\hbar\Omega$. Here, $\mathrm{d} F(t)/\mathrm{d} t=f(t)$ and $k$ is an arbitrary
integer. Note that $F(t)$ satisfies the $\mathcal{T}$-periodicity of the field
since the time average of $f(t)$ vanishes. Thus, $k$ defines a degenerate 
subspace of the extended Hilbert space.  In each degenerate subspace, the 
matrix elements of the perturbation read
\begin{equation}
\label{app:Hperturbation}
({H}_1)_{\alpha\beta} = \frac{1}{\mathcal{T}}\int_0^\mathcal{T} \d t\,
\langle\phi_\alpha^k(t)|H_1|\phi_\beta^k(t)\rangle .
\end{equation}
Therefore, to first order in $H_1/\hbar\Omega$, the Floquet states and the
quasienergies for the Hamiltonian (\ref{app:H}) are obtained by diagonalising
the perturbation matrix \eqref{app:Hperturbation}.

Following \eqref{eq:basetrafo}, the basis states $\vert\varphi_\alpha\rangle$ and 
$\vert\phi_\alpha^k(t)\rangle$ are related by the unitary transformation
\begin{equation}
\label{app:trafo}
 U_0(t) = \exp\left(-\frac{\mathrm{i}}{\hbar}H_0 F(t)\right) 
\end{equation}
as $ \exp\left(-\mathrm{i}k\Omega t\right)\vert\phi_\alpha^k(t)\rangle =
U_0(t)\vert\varphi_\alpha\rangle$.
For the Hamiltonian \eqref{eq:HTLS}, \eqref{app:trafo} corresponds to the 
unitary transformation \eqref{trafo}. 

Within the regime of validity of the perturbative approach, the problem 
is therefore described by the static Hamiltonian
\begin{equation} 
\label{app:Hbar} \bar H_1
= \frac{1}{\mathcal{T}}\int_0^\mathcal{T} \d t\, U_0^\dagger(t)H_1U_0(t).
\end{equation}
Note that after the transformation with \eqref{app:trafo}, the amplitude of
the oscillating part of the new Hamiltonian $U_0^\dagger(t)H_1U_0(t)$ is no 
longer governed by $H_0$, but rather by $H_1$.  Thus, a perturbative treatment 
of the oscillating part is (almost) independent of the original driving amplitude 
in $H_0$.

A particular example for a high-frequency approach along these lines
is a particle moving in a one dimensional 
continuous potential under the influence of a dipole field, i.e. 
$H_1=p^2/2m+V(x)$ and $H_0=\mu x$.  Then, \eqref{app:trafo} constitutes
a gauge transformation and results in a Hamiltonian 
which is again of the form \eqref{app:H}.  A second 
transformation of the type \eqref{app:trafo} yields a periodically 
accelerated potential and defines the so-called Kramers-Henneberger frame 
\cite{Kramers1956a,Henneberger1968a}.

\bibliographystyle{elsart-num}

\begin{thebibliography}{10}
\expandafter\ifx\csname url\endcsname\relax
  \def\url#1{\texttt{#1}}\fi
\expandafter\ifx\csname urlprefix\endcsname\relax\def\urlprefix{URL }\fi

\bibitem{Reed1997a}
M.~A. Reed, C.~Zhou, C.~J. Muller, T.~P. Burgin, J.~M. Tour, Science 278 (1997) 252.

\bibitem{Cui2001a}
X.~D. Cui, A.~Primak, X.~Zarate, J.~Tomfohr, O.~F. Sankey, A.~L. Moore, T.~A.
  Moore, D.~Gust, G.~Harris, S.~M. Lindsay, Science 294 (2001) 571.

\bibitem{Reichert2002a}
J.~Reichert, R.~Ochs, D.~Beckmann, H.~B. Weber, M.~Mayor, H.~von L\"ohneysen,
Phys. Rev. Lett. 88 (2002)
  176804.

\bibitem{Hanggi2002a}
{Special Issue: \emph{Processes in Molecular Wires}, edited by P. H\"anggi, M.
  Ratner, and S. Yaliraki, Chem. Phys. \textbf{281} (2002) 111--502.}

\bibitem{Nitzan2003a}
A.~Nitzan, M.~Ratner, Science 300 (2003) 1384.

\bibitem{Liang2002a}
J.~Park, A. N. Pasupathy, J. I. Goldsmith, C. Chang, Y. Yaish,
J. R. Petta, M. Rinkoski, J. P. Sethna, H. D. Abru\~na, P. L.
McEuen, and  D. C. Ralph, Nature, 417 (2002) 722.

\bibitem{Zhitenev2002a}
N.~B. Zhitenev, H.~Meng, Z.~Bao, Phys. Rev. Lett. 88 (2002) 226801.

\bibitem{Lee2003a}
J.-O. Lee, G.~Lientschnig, F.~Wiertz, M.~Struijk, R.~A.~J. Janssen,
  R.~Egberink, D.~N. Reinhoudt, P.~Hadley, C.~Dekker, Nanoletters 3 (2003) 113.

\bibitem{Lehmann2003a}
J.~Lehmann, S.~Camalet, S.~Kohler, P.~H\"anggi, Chem. Phys. Lett. 368 (2003) 282.

\bibitem{Camalet2003a}
S.~Camalet, J.~Lehmann, S.~Kohler, P.~H\"anggi, Phys. Rev. Lett. 90 (2003) 
210602.

\bibitem{Blick1996a}
R.~H. Blick, R.~J. Haug, J.~Weis, D.~Pfannkuche, K.~von Klitzing, K.~Eberl,
Phys. Rev. B 53 (1996) 7899.

\bibitem{Blanter2000a}
Y.~M. Blanter, M.~B\"uttiker, Phys. Rep. 336 (2000) 1.

\bibitem{Callen1951a}
H.~B. Callen, T.~A. Welton, Phys. Rev. 83 (1951) 34.

\bibitem{Chen1991a}
L.~Y. Chen, C.~S. Ting, Phys. Rev. B 43 (1991) 4534.

\bibitem{Grossmann1991b}
F.~Grossmann, P.~Jung, T.~Dittrich, P.~H\"anggi, Z. Phys. B 84 (1991) 315.

\bibitem{Grossmann1992a}
F.~Gro{\ss}mann, P.~H\"anggi, Europhys. Lett. 18 (1992) 571.

\bibitem{Grossmann1991a}
F.~Grossmann, T.~Dittrich, P.~Jung, P.~H\"anggi, Phys. Rev. Lett. 67 (1991) 516.

\bibitem{Shirley1965a}
J.~H. Shirley, Phys. Rev. 138 (1965) B979.

\bibitem{Sambe1973a}
H.~Sambe, Phys. Rev. A 7 (1973) 2203.

\bibitem{Grifoni1998a}
M.~Grifoni, P.~H\"anggi,  Phys. Rep. 304 (1998) 229.

\bibitem{Kramers1956a}
H.~A. Kramers, Collected Scientific Papers, North Holland, Amsterdam, 1956.

\bibitem{Henneberger1968a}
W.~C. Henneberger, Phys. Rev. Lett. 21 (1968) 838.

\end{thebibliography}

%

\end{document}